
\input vanilla.sty

\headline={\ifnum\pageno=1\firstheadline\else
\ifodd\pageno\rightheadline \else\leftheadline\fi\fi}
\def\firstheadline{\hfil}
\def\rightheadline{\hfil}
\def\leftheadline{\hfil}
	\footline={\ifnum\pageno=1\firstfootline\else\otherfootline\fi}
\def\firstfootline{\rm\hss\folio\hss}
\def\otherfootline{\hfil}

 1
 1
 1

\font\elevenbf=cmbx10 scaled\magstephalf

\font\tenbf=cmbx10
\font\tenrm=cmr10

\font\eightrm=cmr8
\font\eightit=cmti8

\parindent=1.2pc
\magnification=\magstep1
\hsize=6.0truein
\vsize=8.6truein
\nopagenumbers

\centerline{\tenbf WORLD SCIENTIFIC PUBLISHING COMPANY}
\baselineskip=18pt
\centerline{\tenbf PLANCKIAN SCATTERING AND SHOCK WAVE MIXING IN}
\baselineskip=13pt
\centerline{\tenbf GENERAL RELATIVITY AND DILATON GRAVITY}

\centerline{\eightrm PARTHASARATHI MAJUMDAR}
\baselineskip=12pt
\centerline{\eightit The Institute of Mathematical Sciences, CIT
Campus,}
\baselineskip=10pt
\centerline{\eightit Madras 600 113, India.}
\centerline{\eightrm E-mail: partha\@imsc.ernet.in}\vglue0.2cm
\vglue0.6cm
\centerline{\eightrm ABSTRACT}
\vglue0.2cm
{\rightskip=3pc
 \leftskip=3pc
 \eightrm\baselineskip=10pt\noindent
Point particle scattering at Planckian centre-of-mass (cm) energies
and low fixed momentum transfers, occurring due both to
electromagnetic and gravitational interactions, is surveyed, with
particular emphasis on the novel features occurring in electromagnetic
charge-monopole scattering. The issue of possible mixing of the shock
waves occurring due to both kinds of interactions is then addressed
within the framework of Einsteinian general relativity and the
dilatonic extension suggested by string theory.
\vglue0.6cm}
\tenrm\baselineskip=13pt
\leftline{\elevenbf 1. Introduction}
\vglue0.4cm

One of the non-perturbative alternatives commonly pursued in quantum field
theories with
coupling constants generically of order unity or larger, is the semiclassical
approximation.
In this approximation, a conveniently chosen subset of the entire space of
degrees of freedom
is quantized in the classical background provided by the remaining degrees of
freedom. An
oft-quoted example of this approach is the formulation of quantum theories of
`matter' fields
in classical gravitational backgrounds: one starts in this approach with the
assumption that
the quantum
fluctuations of the gravitational background may be ignored. Attempts to go
beyond the
confines of this
approximation would no doubt entail a fully consistent quantum theory of
gravitation which is
still not available. Fortunately there exist kinematical
domains in several situations of physical interest where such an
approximation turns out to be an {\it exact} one, in that,
fluctuations of the non-quantized degrees of freedom essentially
decouple from the physics. In other words, within the kinematical window, the
{\it entire}
dynamics is captured by the semiclassical approximation, and no further
correction is ever
needed. Thus, one can account for all quantum gravitational effects that may
occur within this
kinematical regime, without having (or needing) a full theory of quantum
gravity. Even if we were
to be
able, some day, to formulate such a theory, the physical predictions of this
theory, when
restricted to the appropriate kinematical domain must match the results of the
semiclassical approach. In what follows, we wish to dwell on some instances of
this
fortutious occurrence. We mention in passing that the `solvable' (classical)
part of the
theory still has nontrivial information about interactions in contrast to
standard
perturbation theory which is generally around free field theory.

The kinematical arena for what follows is the large centre-of-mass (cm) energy
($\sqrt{s}$), fixed, small momentum transfer ($\sqrt{t}$) regime; we shall be
interested in the
limiting situation $s/t \rightarrow \infty$. Clearly this corresponds to {\it
almost} forward
scattering at tiny scattering angles and large impact parameters. In this
kinematical regime
there is a decoupling of local field degrees of freedom from the dynamics. The
residual degrees of freedom form a much simpler lower dimensional field theory
which
appears like a boundary field theory. Furthermore, the interplay between
gravitational and
electromagnetic interactions become especially interesting in this
kinematical regime when one of the particles is magnetically charged.
In this case the fine structure constant of electromagnetism $\alpha$
does not evolve with $s$, but increases with
increasing $t$. So if $t$ is held fixed then $\alpha$ does not run at all.
Thus,
in the kinematical region of interest, one expects gravitational
interactions to dominate over electromagnetism. With monopole-charge
scattering, though, this is not the case as we show below.

In this latter situation, with both gravity and electromagnetism
mediated by their respective shock wave fronts, the possibility of an
`interference' of these shock fronts must be taken into account. If
these shock waves do not mix, then earlier results on the
amplitudes for Planckian scattering of point charged particles
(including ours) shall stand vindicated. On the other hand, possible
mixing of these shocks will lead to new physics like a neutral test
particle experiencing effects due to the charge of a boosted charged
black hole. Our analysis, albeit heuristic and hence tentative,
appears to indicate that shock waves do not mix in the case of Einstein
gravity, but they {\it do } for the dilatonic extension currently in
vogue, inspired by string theory.

The survey is organized as follows. In the second section we review
earlier literature$^{1,2}$ on pure electric charge-charge scattering within the
``shock-wave'' picture. Scaling arguments leading to a truncation of the
Maxwell action and eventually to the shock wave picture will be
summarized for completeness. Then we introduce magnetic monopoles in the
theory,
and proceed to generalize the foregoing formalism to calculate the
scattering amplitude. Particular attention will be paid to subtleties
arising from problems like the Dirac string singularity. In the third
section gravity will be introduced and the
interactions involving both electromagnetism and gravity will be
studied. Next we discuss charge-charge and charge-monopole scattering at
Planckian energies. The relative contributions of electromagnetic and
gravitational scattering in the two cases will be contrasted in detail. We
will also comment on
the behavior of singularities, namely the poles in the scattering
amplitude and how they differ from one process to another. In the next
section we address the issue of mixing of shock waves. The
Reissner-Nordstrom metric which describes the static gravitational
field due to a point charge, is boosted to luminal velocities to
determine whether the resulting gravitational shock wave depends on
the charge. We present arguments to the effect that the
charge-dependent piece of the boosted metric can be removed by a
`global' diffeomorphism and hence does not contribute to the shock
wave. Quite a contrasting situation is then shown to exist for a
dilatonic black hole, where the shock wave geometry depends explicitly
on the charge, and hence characterizes a mixing of the two species.
Physical consequences of this mixing are discussed. We end with a few
concluding remarks on future outlook and applications of this approach.
\vglue0.6cm
\leftline{\elevenbf 2. Electromagnetic Scattering at High Energies}
\vglue0.3cm
\leftline{\tenbf 2.1 Effective Theory at High Energies$^2$}
\vglue0.4cm

Suppose there are two spinless
charged particles moving at very high velocities, such that the
center of mass energy $\sqrt s$ is very high. At these energies and very low
squared
momentum transfer $t$, the scattering is almost exclusively in the forward
direction, with a very large impact parameter. This small ratio between the
scales of momenta in the transverse and longitudinal directions enables us
to associate two widely different length scales with the longitudinal and
transverse directions, Thus, we scale the null coordinates $x^{\pm}$
such that $x^\alpha \rightarrow \lambda
x^\alpha$ and $x^{i} \rightarrow x^{i}$, where
$\alpha$ runs over the
light cone indices $+,-$, while $i$ signifies the transverse
coordinates $x,y$. Note that $\lambda$ is not a new parameter in the theory; it
merely signifies the large kinematical ratio under consideration.
Under this scaling the $A_{\mu}~s$ transform as
$A_\alpha \rightarrow
{\lambda}^{-1}A_{\alpha}$. The transverse $A_i~s$ remain unchanged.
The transformed action now has the form
$$
S~=~-{1 \over 4}\int d^4x \left( {\lambda}^{-2}F_{\alpha \beta}F^{\alpha
\beta}~+~2F_{\alpha i}F^{\alpha i}~+~
{\lambda}^2 F_{ij}F^{ij} \right) ~ \eqno(1).
$$
The parameter $\lambda$ may now be chosen to depend on $s$ :
$$
\lambda = {k \over {\sqrt s}}\rightarrow 0~~,
$$
where $k$ is a finite constant having dimensions of energy.
Then the limit $s \rightarrow \infty$ becomes equivalent to the
limit $\lambda \rightarrow 0$. Thus in this kinematical regime,
the transverse part of the action with $F_{ij}$ can be ignored and what we have
left is an effective action of the form
$$
S~=~-{1 \over 4}\int d^4x \left( {\lambda}^{-2}F_{\alpha \beta}F^{\alpha
\beta}~+~2F_{\alpha i}F^{\alpha i}\right)~~.
$$
Notice that in the partition function the fluctuations of the
term $F_{\alpha \beta}F^{\alpha \beta}$ are suppressed
in the imaginary exponent (due to
the smallness of $\lambda$) and the configuration with the dominant
contribution is
$F_{\alpha \beta}=0$ i.e. $F^{+-}=E_z=0$ $^4$.
This shows that the electric field is localized in the transverse
plane. Similarly, if we write the original action in the dual
formalism, with the $F_{\mu \nu} \rightarrow {\tilde F}_{\mu \nu}$, then
${\tilde F}^{+-}=B_z=0$. This brings us to the {\it shock wave picture}:
fields due to processes characterized by longitudinal momenta that are
overwhelmingly larger than transverse momenta are essentially
confined to the plane (called the `shock front')
perpendicular to the direction of motion of the source particles. Thus, if we
were to describe
the gauge field interaction in terms of currents $j^{\mu}$, then only the light
cone
components $j^{\pm}$ would be physically relevant. Furthermore, if these
currents were to be associated with charges moving almost
luminally, then
 $$
j_{\pm}~=~j_{\pm} \left( x^{\pm}, {\vec r}_{\perp} \right) ~~~~
j^{i}(x)~=~0
$$
This allows us to define two functions $k^{+}$ and $k^{-}$, where
$$\eqalign{
j_{+} &~=~{\partial}_{-} k^{-} \left( x^{-}, {\vec r}_{\perp} \right) \cr
j_{-}&~=~{\partial}_{+} k^{+} \left( x^{+}, {\vec r}_{\perp} \right) }
\eqno(2)$$

In short, if we define a vector $k$ such that
$$
k\left(x\right)~=~k^{+}\left(x^{+}, {\vec r}_{\perp} \right) -
k^{-}\left(x^{-}, {\vec r}_{\perp} \right)
$$
then
$$
j^{\alpha}~=~{\epsilon}^{\alpha \beta}{\partial}_{\beta}k. $$

The flatness condition $F^{+-}=0$ above admits a solution in terms of the
light cone components of the gauge
potential $A_{\pm}={\partial}_{\pm}{\Omega}$. If, further, we impose
the Landau gauge ${\partial}_{\mu}A^{\mu}=0$, $\Omega$ obeys d`Alembert's
equation
$$
{\partial}_{+}{\partial}_{-}{\Omega}~=~0
$$
which implies
$$
\Omega~=~{\Omega}^{+}\left(x^{+}, {\vec r}_{\perp} \right) +
{\Omega}^{-}\left(x^{-}, {\vec r}_{\perp} \right).
$$
It is then easy to show that the electromagnetic
Lagrange density
$${\cal L}~=~ -{1 \over 4}F_{\mu \nu}F^{\mu
\nu} - j^{\mu}A_{\mu}$$
can be written as
$$
{\cal L}~=~-{1 \over 2}{\partial}_{-}{\Omega}^{-}{\vec
\nabla}^2{\partial}_{+}{\Omega}^{+}~-~
{1 \over 2}{\partial}_{+}{\Omega}^{+}{\vec
\nabla}^2{\partial}_{-}{\Omega}^{-}
- \partial_+k^+ \partial_-\Omega^-
- \partial_-k^- \partial_+\Omega^+
$$
which reduces to a total derivative  in the light cone coordinates
$$
{\cal L}~=~-{\partial}_{-} \left( {1 \over 2}{\Omega}^{-}~{\vec
\nabla}^2 {\partial}_{+}{\Omega}^{+} +
{\partial}_{+}k^{+}~{\Omega}^{-} \right ) - {\partial}_{+} \left( {1
\over 2} {\Omega}^{+}~{\vec \nabla}^2 {\partial}_{-} {\Omega}^{-}
+ {\partial}_{-}k^{-} {\Omega}^{+} \right).
$$
This shows that the action $S=\int d^4 x {\cal L}$ is a surface term
defined on the boundary of null plane:
$$
S~=~\oint~d\tau \int~d^2r_{\perp}~ \left( {1 \over 2}
{\Omega}^{-}~{\vec \nabla}^2 {\dot {\Omega}}^{+} - {1 \over
2}{\Omega}^{+}~{\vec \nabla}^2{\dot {\Omega}}^{-} + {\dot k}^{+}
{\Omega}^{-} - {\dot k}^{-} {\Omega}^{+} \right) . \eqno(3)
$$
Here all the quantities are evaluated on the boundary contour
parametrized by the affine parameter $\tau$. An overdot
denotes ${\partial}/{\partial}\tau$. Thus $\Omega$s are the only surviving
dynamical degrees of freedom in the problem. This simplification
of the action has its origin in the kinematics of the situation. One
can show that the shock
front discussed earlier emerges as the solution of the classical
equations of motion for the above action.
\vglue .4cm
\leftline{\tenbf 2.2 Charge-charge scattering$^2$}
\vglue .4cm

The foregoing analysis allows us to compute {\it exactly} the $S$-matrix
for the scattering of two highly energetic particles assumed to  carry
electric charge.  Making use of Lorentz covariance of the theory, we
will do the calculations in a special inertial frame in which one
of the charges moves with velocity close to luminal, while the other is
moving relatively slowly. The shock wave front due to the former extends
over the entire transverse plane. Thus, the target particle, assumed to be
moving in a direction opposite to that of the source, encounters this
shock wave and its wave function acquires an Aharanov-Bohm type phase
factor. The overlap between the wave functions of the target particle
before and after its encounter with the shock front leads to the
scattering amplitude.

In the limit $\beta \rightarrow 1$, the potential of the ultrarelativistic
`source'
particle is found to be a pure gauge almost everywhere except on
the shock plane where it has a discontinuity,
$$\eqalign{
\tilde A^{0}&= \tilde A^{z}=-2e'~ \ln (\mu r_{\perp})~ \delta(x^{-})
, \cr
{\tilde A}_{\perp}^{i}&= 0,~~~~~~~~ i=1,2. }
\eqno(4)$$
Here, $\mu$ is a dimensional parameter inserted to make the logarithm in
(4) dimensionless. The potential $\tilde A^{\mu}$ is singular on
the shock
plane $(x^{-}=0)$ and is gauge equivalent to the potential
$A^{'}{}^{\mu}$ where $A^{'}{}^{\mu}=\tilde A^{\mu}+\partial ^{\mu}\Lambda,
{}~\Lambda $ being a Lorentz scalar. Choosing $\Lambda$ to be
$-2e' \theta(x^-) \ln \mu r_{\perp}~ $, we get
$$
A^{'}{}^{0}=A^{'}{}^{3}=0,~~~~~
\vec{A}^{'}{}_{\perp}=-2e^{'}\theta (x^{-})
\vec \nabla \ln \mu r_{\perp} \eqno(5)
$$
We see that the gauged transformed vector potential is a pure
gauge everywhere except on the hyperplane $x^{-}=0$ which is also
the shock plane. Thus as one expects, the fields are
non-vanishing only on this plane and are given by
$$\eqalign{
E^i &=  {2e^{'} r_{\perp}^i \over {r_{\perp}^2}}{\delta
(x^{-})},~~~~~~~~~~ E^z=0 \cr
B^i &= -{2e^{'} \epsilon_{ij} r_{\perp}^j \over
r_{\perp}^2}{\delta
 (x^{-})} ,~~~~ B^z=0~.} \eqno(6) $$
These singular field configurations cause an instantaneous interaction
with the (slower) target particle. We now proceed to calculate the results of
such
interaction$^3$.

Consider the wave function of the slow particle; for early times $t<z$ the
particle is free and its wavefunction is just a plane wave given by,
$$
\psi _<~(x^{\pm}, {\vec r}_{\perp}) =~\psi _0~=~\exp[ipx]~~~~for~ x^-<0~.
$$
with momentum eigenvalue $p^{\mu}$.
Immediately after the shock front passes by, its interaction with
the gauge potential enters via the `minimal coupling
prescription' by which we replace all the $\partial_{\mu}$'s by
$\partial_{\mu}-ieA_{\mu}$. The corresponding wavefunction
acquires a multiplicative phase factor $exp\left(ie\int dx^{\mu}A_{\mu}
\right)$.
Thus from equation (5), for $x^{-}>0$,
the modified wavefunction is
$$
\psi _>(x^{\pm}, {\vec r}_{\perp})~ =~ \exp~[-iee^{'}
\ln(\mu^{2}r^{2}_{\perp})]~\psi_0
{}~~~~for~ x^->0
\eqno(7)
$$
The wavefunction $\psi_>$ can now be expanded in terms of
the complete set of
momentum eigenfunctions (plane waves) with suitable coefficients
in the following form
$$
\psi _> = \int dk_+d^2k_{\bot}~A(k_+ , {\vec k}_{\bot}
)\exp[i{\vec k}_{\bot} \cdot {\vec r}_\bot -ik_+x^- -
ik_-x^+ ] \eqno(8)
$$
with the on shell condition $k_+ = {(k^2_{\perp} +
m^2)/k_-}~.$
Here,
$$
A(k_+,k_\bot) = {\delta (k_+ - p_+)
\over (2 \pi)^2}{\int
d^2r_{\bot} \exp i \left(-2ee^{'}~\ln(\mu r_\perp) +
{\vec q} \cdot {\vec r}_\bot \right)}~,
$$
where ${\vec q} \equiv
{\vec p}_{\bot} - {\vec
 k}_{\bot}$ is the
transverse momentum
transfer, $k$ and $p$ being the final and initial momenta
respectively.  The integration over the transverse $x-y$ plane can be
performed exactly$^3$ yielding the amplitude
$$
f(s,t)= {k_+ \over 4\pi k_0}~{\delta(k_+ - p_+)}
{\Gamma
(1-iee') \over \Gamma (
iee')}\left (4 \over
-t \right
 )^{1-iee'}~~.
\eqno(9) $$
where we have put in the canonical kinematical factors. $t \equiv
-q^2$ is the transverse momentum transfer. With this
amplitude, one can easily show that the scattering cross section is
$$
{d^2\sigma \over d\vec k_{\perp}^2}~ \sim~ {(ee^{'})^2 \over t^2 }~~,
$$
where we have used a property of the gamma function, namely
${|\Gamma (a+ib)|}={|\Gamma (a-ib)|}$, $a$ and $b$ being real.

It has been shown$^2$ that this scattering amplitude is
identical to the amplitude obtained in the Eikonal approximation where
virtual momenta of exchanged quanta are ignored in comparison to external
momenta, leading to a resummation of a class of Feynman graphs. Since,
generically $ee'={\cal O}({1 \over 137})$, this
approximation will receive usual perturbative radiative corrections.
The second order pole singularity in
the cross section as $t \rightarrow 0$ is, of course, typical of processes
where massless quanta are exchanged.
\vglue 0.4cm
\leftline{\tenbf 2.3 Charge - monopole scattering$^4$}
\vglue .4cm

Now that we have calculated the amplitude of the
scattering of two charges, one can inquire as to  what
changes, if any, will take place if we replace one of the charges
by a Dirac magnetic monopole. This question is worth pursuing for various
reasons. First of all, the (albeit imagined) existence of monopoles will
imply that the Maxwell equations assume a more symmetric form, due to the
property of duality of field strengths and electric and magnetic charges.
Within quantum mechanics, as Dirac has shown, monopoles offer a
unique explanation of the quantized nature of electric
charge. But as is well-known, introduction of monopoles in the
theory brings in other problems such as singularities in the
vector potential. It will be interesting to see how one can deal
with them in the present formalism and investigate the range of validity
of the shock wave picture in this context. One should also keep in mind
the fact that a satisfactory local quantum field theory for monopoles is
still lacking. Further, given Dirac's quantization condition, monopole
electrodynamics cannot be understood in perturbative terms around some
non-interacting situation. Thus, as advertized earlier, the shock wave
picture may be one of the few important probes available for such processes.

Recall, however, it is not possible to choose a single
non-singular potential to describe the field of the monopole everywhere.
We need at least two such
potentials, each being well behaved in some region and being related by a
local gauge transformation in the overlapping region.  In spherical polar
coordinates, these potentials can be chosen as
$$
\eqalign{
{\vec {A}}^I ~ & = ~ {g \over {r \sin \theta} } (1 - \cos
\theta) {\hat \phi}~,~~~~~  0 \leq \theta < \pi \cr
{\vec{A}}~^{II}~ & = ~{-~g \over { r \sin \theta}}
(1+\cos\theta){\hat \phi}~.~~~~~  0 < \theta \leq
\pi~.} \eqno(10) $$
The Dirac strings associated with the two potentials are along
the semi infinite lines $\theta= \pi$ and $0$ respectively,
i.e. along the negative and positive halves of the $z$
axis. ${\vec A}^{I}$ and ${\vec A}^{II}$ become singular along these two
lines respectively.
It may be noted that here we have made the gauge choice
$A^0=0$, and have chosen an orientation of our coordinates such
that only the $x$ and $y$
components survive. In the region $-\pi < \phi < \pi$, where either of
$\vec {A}~^{I}$ or $\vec {A}~^{II}$ may be used, they are related
by a gauge transformation with the gauge parameter $2g\phi$. It
can be readily verified that
$$
\vec \nabla  \times \vec{A}~^{I} ~=~ \vec \nabla \times
\vec{A}~^{II} ~=~{g \over r^2}\hat{r}.
$$
Here the curls are taken in the respective regions of
validity of the potentials. In the following calculations for
convenience we shall work with ${\vec A}~^{I}$ only,
but all subsequent results will be independent of this particular choice.

As in the last section, we give the monopole a Lorentz
boost of magnitude $\beta$ along the positive z axis.
It can be shown that
if equations (10) are rewritten in cartesian coordinates,
then
$\vec{A}~^{I}$ transforms to
$$
{}^{\beta}{A}^I_i={-g \epsilon_{ij} r^j_{\perp} \over
r^2_{\perp}
 }\left [1 - {{z
 - \beta t} \over R_{\beta}} \right ] ~. \eqno(11)
$$
Before proceeding further, let us examine the behavior of the Dirac
strings under Lorentz boosts. For this purpose it is convenient to write
equation (11) in the following form.
$$
^{\beta}{\vec A}^{I}~=~{g \over r_{\bot}} \left[ 1 - {{z - \beta t}
\over R_{\beta}} \right] \hat \phi
\eqno(12)
$$
On the $z$- axis $\left( \theta=0~~ or~~ \pi \right)$, we have $ r_{\bot}
\rightarrow 0$ implying that $R_{\beta} \rightarrow |z - \beta t|$. Thus
the above equation reduces to
$$
^{\beta}{\vec A}^{I}~=~{g \over r_{\bot}} \left[ 1 - sgn (z - \beta t)
\right] .
$$
Thus for $z>\beta t$, i.e. in front of the boosted monopole the vector
potential vanishes, while it becomes singular behind it $\left( z <
\beta t \right)$. It is as if the monopole drags the Dirac string along
with it and as in the static case, the semi - infinite line of
singularity originates from it. Similarly, by looking at the boosted
potential $\vec A^{II}$, it can be easily verified that for this, the
string is always in front of the monopole and `pushed' by it as it moves.
These results also hold in the limit $\beta \rightarrow 1$, i.e. for the
potential,
$$
\vec{\tilde A}~^I_0~\equiv \lim_{\beta \rightarrow 1}~
^{\beta}{\vec A}_i~^I = {2g \over r_{\perp}}~\theta(x^-) {\hat \phi}.
\eqno(13) $$
The corresponding electromagnetic fields are
$$\eqalign{
B^i ~ & =~  {2g r_{\perp}^i \over {r_{\perp}^2}}{\delta
(x^{-})},~~B^z=0  \cr
E^i ~ &=~ {2g \epsilon_{ij} r_{\perp}^j \over
r_{\perp}^2}{\delta
(x^{-})} ,~~ E^z=0~. }\eqno(14) $$
Unlike the fields of a charge in motion, here the magnetic field
is radial, whereas the electric field is circular on the shock
plane. Here also $\vec A^{I}_{0}$ is a pure gauge everywhere except on
the null plane $x^{-}=0$. It may be noted that the above $\vec E$
and $\vec B$ fields can be obtained by making the following
transformations in (14):
$e^{'} \rightarrow g$, $\vec E
\rightarrow \vec B$ and $\vec B \rightarrow -\vec E$. This is a
consequence of the duality symmetry in Maxwell's equations
incorporating monopoles.

To compute the scattering amplitude, we first rewrite
${\vec {\tilde A}}~^I_0$ in (12) as a total derivative in the
following form
$$
\vec {\tilde A}~^I_0~ =~ 2g \theta (x^{-}) \vec {\nabla} \phi ~.
\eqno(15) $$
We note in passing that the gauge potentials for a luminally boosted
electric charge (13) and monopole (15), both given
as total derivatives on the transverse plane, form the real and imaginary
parts respectively of the gradient of the holomorphic function $lnz$
where $z \equiv r_{\perp} e^{i \phi}$, where $\phi$ is now the azimuthal
angle on the transverse plane.

For $t<z$, i.e. before the arrival  of the monopole with its shock front, the
wave function of the charge $e$ is once again the plane wave
$$
\psi _<(x^{\pm},{\vec r}_{\perp})~ =~\psi _0 ~~~~for~ x^-<0~. \eqno(16)
$$
After encountering the shock wave, it is modified by the
gauge potential dependent phase factor. The final form of the
wave function is
$$
\psi _>(x^{\pm},{\vec r}_{\perp})~ =~ exp[i2eg \phi]~\psi '_0  ~~~~for~ x^->0
$$
by virtue of the potential (15) with the usual
requirement of continuity. At this point we make the additional
assumption of Dirac
quantization namely, for an interacting monopole-charge
system, the magnitudes of their electric and magnetic charge must be
constrained by the relation
$$
e~g~=~{n \over 2},~~~~~n~=~0,\pm1,\pm2,....
$$
Thus we get
$$
\psi_>~=~e^{in\phi}{\psi}^{'}_{0}~.
$$
This sort of phase factor in the small angle scattering of a
monopole and a charge was first found by Goldhaber$^5$ using more
standard techniques.
Expanding $\psi_>$ in plane waves as before we get an integral
expression for the scattering amplitude as follows
$$
A(k_+,k_\bot) = {\delta (k_+ - p_+) \over (2 \pi)^2}{\int
d^2r_{\bot} \exp i\left(n\phi + {\vec q} \cdot {\vec r}_\bot
\right)}~. \eqno(17)
$$
Once again $\vec q \equiv p_\perp - k_\perp$ is the momentum
transfer and as before we have the dispersion relation $k_+ =
{(k^2_{\bot} + m^2)/k_-}~.$
By conveniently choosing the orientation of the transverse
axes as in the previous section, the angular integration
gives
$(1 / q^2) \int_{0}^{\infty} d\rho ~ \rho J_n(\rho)~,$
where $J_n(\rho)$ is the Bessel function of order $n$. This
integral is also standard and the result is
$$
\left ({1\over -t}\right )  {2\Gamma (1+ {n \over 2}) \over
\Gamma ({n \over 2})} ~, \eqno(18)
$$
Here we note an important difference with the previously
calculated charge-charge amplitude. There the arguments of the
gamma functions were complex, whereas in this case they are
real. In fact, the amplitude in this case is simply
$$
f(s,t)~=~{k_+ \over 2\pi k_0}~{\delta (k_+ - p_+) \left ({n \over
-t}\right)}~, \eqno(19)
$$
where we have incorporated the canonical kinematical factors.
Such factorization makes the expression for the amplitude simple. We
observe that it is proportional to the
monopole strength $n$. It follows that the scattering cross
section becomes
$$
{d^2\sigma \over d\vec k_{\perp}^2}~ \sim~ {n^2 \over t^2} \eqno(20)
$$

It may be mentioned that we would have obtained the same result if we
had used the second of the gauge potentials in
(13) and performed the Lorentz boost etc.  One way to see this
is by noting that the potentials, boosted to $\beta \approx 1$, are both
gauge
equivalent to a gauge potential $A'_{\mu}$ given by
$$
{\vec A}'_{\perp}~=~0~=~A'_+~;~~~A'_-~=~2g \phi \delta(x^-)~~everywhere ~.
\eqno(21)
$$
The apparent disappearance of the Dirac string singularity in this
gauge is a red herring;  the gauge transformation has flipped the
Dirac string onto the shock plane, thus preventing it from being
manifest. More importantly, the gauge potential, though globally defined
functionally, is not single-valued, being a monotonic function of a
periodic angular variable. Thus, the singularity has been traded in for
non-single-valuedness. Of course, the theory
of fields which are not single-valued functions is in no way easier to
formulate
than that for singular fields. It is interesting to note further  that
for boost velocities that are subluminal, one
cannot obtain a globally defined potential $A'^{\mu}$ in any gauge.

We would like to make a few more remarks at this point.
First of all, if we choose another Lorentz frame in which the electric
charge is lightlike while the monopole is moving slowly,
we would get
identical results from the dual formalism wherein one introduces a gauge
potential
$A_{\mu}^M$
such that the dual field strength ${\tilde F_{\mu \nu}} \equiv \partial_{[\mu}
A_{\nu]}^M $.
If this gauge potential is used to
define electric and magnetic fields, the standard field tensor $F^{\mu \nu}$
must satisfy a
Bianchi identity of the form $\partial_{\mu} F^{\mu\nu}=0$
which would then imply that
the gauge potential due to a point charge must have a Dirac
string singularity. Further, the monopole will behave
identically to
the point charge of the usual formalism, so that our method
above is
readily adapted to produce identical consequences. Second, one
can also
treat the scattering of two Dirac monopoles in the same
kinematical limit exactly as in subsection 2.1, using this dual formalism. This
would yield a
result identical to the one for the electric charge case,
with $e$ and $e'$ being replaced by $g$ and $g'$, the monopole
charges. Finally, having dealt with particles carrying either electric
or magnetic charge, it is straightforward to extend our
calculations when one of them is a dyon, that is, it has
both electric and  magnetic charge. The
electromagnetic fields on the
shock front of the
boosted dyon will be the superposition of the fields
produced by a fast charge and a monopole.
Also, depending upon
the nature of the charge on the other particle (electric or
magnetic), one must employ the usual or the dual formalism.
\vglue.4cm
\leftline{\tenbf 2.4 Dyon - dyon scattering$^6$ }
\vglue.4cm

With the above observations we are in a position
to address the problem of dyon-dyon scattering in this
formalism. Consider two dyons $(e_1,g_1)$ and
$(e_2,g_2)$, where the ordered pair denotes its electric and
magnetic charge contents respectively. Let us assume that the first one
is ultra relativistic. By means of an electromagnetic duality
transformation we can `rotate' the dyon $(e_1, g_1)$ by an angle
$\theta$, so that
the new values of electric and magnetic charges become $e'$ and $g'$.
The same duality transformation rotates the second dyon to $(e,g)$.
Since physical observables do not depend on the parameter $\theta$, we
can make use of this symmetry and choose it to be such that
$$\tan \theta = {g_2 \over e_2}~. \eqno(22) $$
This implies that the first dyon transforms to
$$\eqalign{
e'&={e_1e_2 + g_1g_2 \over \sqrt {e_2^2 + g_2^2}}  \cr
g'&={-e_1g_2 + g_1e_2 \over \sqrt {e_2^2 + g_2^2}} } \eqno(23) $$
while for the second dyon
$$\eqalign{
e&=\sqrt {e_2^2 + g_2^2}  \cr
g&=0 .} \eqno(24) $$
This shows that the slow test dyon has been rotated to a pure electric
charge. Then from the results derived previously, the total phase shift
in its wavefunction after being hit by the shock wave of the dyon
$\left( e', g' \right)$ is $ \left[ ee'
\ln {\mu}^2{r_\bot}^2 + 2eg'
\phi \right] .  $
Having found this, we can express this in terms of the parameters of the
two dyons we started with. The result is $ \left[ \left( e_1e_2 + g_1g_2
\right) \ln {\mu}^2{r_\bot}^2  - 2\left( e_1g_2 - g_1e_2 \right) \phi
\right] .$
The calculation of the scattering amplitude now
becomes straightforward. It may be noted that the quantities
$ \left( e_1e_2 + g_1g_2
\right)$ and $\left( e_1g_2 - g_1e_2 \right)$ are the only combinations
of the electric charges $e_1,e_2$ and the magnetic charges $g_1,g_2$
that are invariant under duality rotations$^7$ . Thus it is
remarkable that the total phase shift and hence the scattering amplitude
depends only on these combinations. Alternatively, we could also have
made the choice  $ \tan \theta = -e_2 / g_2 $, in which case $e$ would
become zero and the second dyon transforms into a monopole. Obviously
these different choices are merely for convenience and the scattering
amplitude does not depend on it. Thus dyon-dyon scattering can always be
reduced to dyon-charge or dyon-monopole scattering. Also note that by a
duality rotation the
usual Dirac quantization condition gets transformed into the generalized
expression
$$ e_1g_2 - e_2g_1 = {n \over 2} \eqno(25)  .$$
This implies that the
second term in the phase shift becomes $n \phi$ as in the
charge-monopole scattering case.

Finally, we can ask the question as to what happens if we consider a
massive vector field e.g. that described by the Proca Lagrangian
$${\cal L}~=~-{1 \over 4}F_{\mu \nu}F^{\mu \nu} + {{\mu}^2 \over 2}
A_{\mu}A^{\mu} \eqno(26)$$
The solution in the static limit for $A^{\mu}$ in the Lorentz gauge is
given by
$$A^{0}~=~{e'\exp\left(-\mu r\right) \over r}~,~~~~~~~A^{i}~=~0,
\eqno(27) $$
where $e'$ is a point charge at rest. Formally we can apply a Lorentz
boost to this potential and try to take the limit $\beta \rightarrow 1$.
The result is
$${}^{\beta}A^{\mu}~=~{\eta}^{\mu}{e'\exp \left(-\mu R_{\beta}/{\sqrt
{1 - {\beta}^2}}\right) \over R_{\beta}} \eqno(28)$$
which vanishes identicaly when we take the limit $\beta \rightarrow 1$.
Thus no shock wave emerges in this case and there are no
$\delta$- function electromagnetic fields on the null plane $x^-=0$. This
observation can also be understood as follows. In the formulation of the
boundary field theory in section 2.1 it was shown
that the gauge parameter $\Omega \left({\Omega}^+,{\Omega}^-\right)$ was
the only dynamical degree of freedom in the theory and the corresponding
equations of motion yielded the shock wave picture. On the other hand,
the Lagrangian of the massive vector field does not have the required
gauge invariant structure to admit of such a parameter.
This accounts for the absence of the shock wave.
\vglue0.6cm
\leftline{\elevenbf 3. Electromagnetic vs Gravitational Scattering at
Planckian Energies} \vglue0.4cm
\leftline{\tenbf 3.1 Spacetime around a massless particle}
\vglue .4cm

At Planckian cm energies the Einstein action also undergoes a
truncation akin to the electromagnetic situation. The spacetime
geometry that emerges for a particle boosted to velocities
close to luminal, is expected to emerge from the coupling of the above
truncated action to a suitably constrained matter energy-momentum tensor.
This has been done in ref.[1]. Identical answers can however be
obtained by a process of {\it boosting} the static (Schwarzschild) metric
due to
a point particle, adopted in ref.[3]; we sketch this approach
below. Essentially this boosting means the  mapping of a solution of
Einstein equation with a massless particle, to Minkowski space but
with one of
the null coordinates shifted non-trivially, now without any massless
particle present$^9$. It is argued below how this can be
interpreted as a gravitational shock wave.

Once again we choose to carry out
the analysis in a Lorentz
frame in which the velocity of one particle is very much greater than
that of the other. We know that the space time around a point
particle is spherically symmetric and is described by what is
known as the Schwarzchild metric.
If we assume the mass $m$ of the particle to be small, then it is
given in the Minkowski coordinates ($T,x,y,Z$) by,
$$
ds^2~=~-\left(1-{2Gm \over R}\right)~dT^2 + \left(1+{2Gm \over R}
\right)
{}~\left(dx^2+dy^2+dZ^2 \right).
\eqno(29)
$$
where $R=\sqrt {x^2+y^2+Z^2}$ and $m \ll R/G$ $^5$.
If the above coordinate
system is moving with a relative velocity $\beta$ with respect to
coordinates $\left(t,x,y,z \right)$ then the two are related by a
Lorentz transformation of the form
$$\eqalign{
T~&=~t\cosh \theta - z\sinh \theta,     \cr
Z~&=~-t\sinh \theta+z\cosh\theta,
} \eqno(30) $$
$\theta$ is called the rapidity which is related to the
boost velocity by the relation
$$
\tanh \theta~=~\beta.
$$
Now to take the limit $\beta \rightarrow
1$ or alternatively $\theta \rightarrow \infty$, we also set
$$
m~=~2p_{0}~e^{-\theta},
$$
where the rest energy of the particle is $2p_{0}>0$.
This parametrization is consistent with the fact that
the mass of the particle must exponentially vanish as its velocity
approaches that of light. When we substitute equation (30)
in equation~(29), we have the metric due to a
particle moving at the speed of light in the $x^{+}$-direction
(i.e along $x^{-}=0$). In terms of the lightcone and the
transverse coordinates this metric becomes
$$
ds^2~=~\left(1+{2Gm \over R} \right)\left[-dx^{-}~dx^{+} + dx^2
+dy^2\right] + {4Gm \over R}\left[{p_{0}  \over m}dx{-} +
{m \over {4p_{0}}}dx^{+}
\right]^2,
$$
with
$$
R^2~=~x^2 + y^2 + \left( {p_{0} \over m}x^{-} - {m \over {4p_{0}}}x^{+}
 \right)^2. \eqno(31)
$$
Using this and neglecting terms of order $m$ or above, we
get the limiting form of the metric
$$
\lim_{m \rightarrow 0}~ds^2= -dx^{-} \left(dx^{+} -
4Gp_{0} {dx^{-} \over
|x^{-}|} \right ) + dx^2 +dy^2 ,
$$
where the limit is evaluated at $x^{-} \neq 0$ and ($x^{+},x,y$) fixed.
Defining a new set
of coordinates through the relation
$$\eqalign{
dx'^{+}&~=~dx^{+}-
{4Gp_{0}~dx^{-} \over
|x^{-}|},  \cr
dx'^-&~=~dx^-  \cr
dx'^i&~=~dx^i ~, } \eqno(32) $$
we observe that the above metric is just a flat Minkowski
metric
$$
ds^2 = -dx'^{-}~dx'^{+} + dx'^2 + dy'^2.  \eqno(33)
$$
The crucial point to note here is that the metric suffers a
discontinuity at $x^{-}=0$ through the term ${|x^{-}|}^{-1}$.
Now, taking the leading order terms in equation
(33), it can be shown that $dx^{-}/x^{-}=dR/R$, which gives
$$
dx'^{+}=dx^{+}- \theta (x^{-})~{4Gp_{0}~dR \over R}
$$
A solution of the above equation near the null plane ( $|x^-|
\rightarrow 0$ ) is,
$$
x'^+~=~x^{+} + 2Gp_{0}~ \theta (x^{-})
\ln \left( {\mu}^2 {r_\perp}^2 \right) .
\eqno(34)
$$
Note that the coordinates $x^-$ and $x^i$ remain unchanged.
This step function at the null plane $x^{-}=0$ is the gravitational
equivalent of the electromagnetic shock-wave. There we had a
similar discontinuity in the gauge potential $A^{\mu}$. Here
we have two flat regions of space-time
corresponding to $t<z$ and $t>z$ which are glued together at the
null plane $t=z$ (or~$x^{-}=0$). However there is a shift of
coordinates at this plane given by equation (32). It is as
if a two dimensional flat space-time on the $t-z$ plane is cut
along the line $t=z$ and pasted back again after being shifted along this
line by the amount given above.

Now that we have found the metric around a lightlike particle, in
principle we should be able to predict the behavior of another
(slower) test particle encountering it. Since the
sole effect of the gravitational shock wave is the cutting and pasting of
the Minkowski space along the null direction $x^-=0$ after a shift of the
$x^+$ coordinate, it is easy to see that the test particle wave function
will acquire a phase factor upon passing through this shock front. One
more remark is in order at this point. The
logarithmic singularity in the expression for the shift in the coordinate
$x^+$ in equation (32) causes an infinite time delay of all
interactions via virtual particle exchanges. This shows that it is the
shock wave interactions which dominate over all standard field theoretic
effects like particle creation via brehmstrahlung etc. However, as we
shall show later, the gravitational shock wave may not dominate in all
situations where other interactions mediated also by shock wavefronts exist.
\vglue .4cm
\leftline{\tenbf 3.2 Gravitational scattering}
\vglue .4cm

To begin with we will assume the particles to be neutral and as
before, also spinless. We look at the behavior
of the wavefunction of a slow test
particle in the background metric of the lightlike particle
carrying with it a `gravitational' shock wave.
Before the arrival of the shock wave ($x^{-}<0$),
the test particle is in a
flat space time as derived in the last section. Thus, as before,
 its quantum
mechanical wave function is a plane wave of the form
$$
\psi_<~(x^{\pm}, {\vec r}_{\perp})~=~e^{ipx}
$$
with definite momentum $p^{\mu}$.
This can be written in terms of the lightcone and transverse
coordinates as
$$
\psi_<(x^{\pm}, r_{\perp})~
=~\exp i \left[ p_{\perp} x_{\perp} - p_{+}x^{-} - p_{-}x^{+}
\right ]
$$
On encountering the shock wave, it is transported to another
flat space time defined by $x^{-}>0$ which is related to the
previous one by a shift in the $x^{+}$ coordinates.
{}From the explicit expression for this shift in equation
(32) we see that the wavefunction immediately gets modified into
$$
\psi_>(x^{\pm},
{\vec r}_{\perp})~=~\exp i \left[ p_{\perp} x_{\perp} - p_{-} \left( x^{+} +
2G p_0 \ln {r_\perp}^2 \right) \right],
$$
which is also a plane wave but in the new coordinates.
We have put $\mu = 1$ in equation
(32) and evaluated the above
at $x^{-}=0^{+}$.
Noting that the factor $2G p_{-} p_{0}$
can be written as $Gs$, the phase shift in the final
wave function is
$-Gs \ln r^2_{\perp}.$
But this is just
the electromagnetic phase shift that we got in the last section
in the case of charge-charge scattering with $Gs$ replacing
the earlier coupling $ee'$. This implies that
the scattering amplitude will
also be the same as the previous case with this replacement.
Consequently we have for the gravitational scattering of the two
particles,
$$
f(s,t)= {k_+ \over 4\pi k_0}~{\delta(k_+ - p_+)}
{\Gamma
(1-iGs) \over \Gamma (
iGs)}\left (4 \over
-t \right
 )^{1-iGs}~~. \eqno(35)
$$
The corresponding differential cross section is
$$
{d^2\sigma \over d\vec k_{\perp}^2}~ \sim~ {G^2s^2 \over
t^2}
$$
Despite the striking similarity with electromagnetism, there is an
important difference here. The coupling is now proportional to $s$, the
square of the center of mass energy. The above cross section seems to
increase without limit
with increase of $s$, thus violating unitarity. To understand this, we must
note that at super-Planckian energies one expects gravitational collapse
and inelastic processes to take place. Hence the
above expression fails to be a faithful representation of the
actual scattering and one has to invoke a full
theory of quantum gravity at such extreme energies$^3$.
Similar arguments hold good for all the other cross
sections found in this paper.

Another important point to note is the structure of poles in the
scattering amplitude (35). It seems that there is a `bound
state'  spectrum at
$$ Gs = -iN ~~,~~~~~ N = 1,2,3,\ldots~~~~~~.$$
It has been remarked in ref.[6] that the $t$- dependence of the
residues of the poles can be expressed as polynomials in $t$ with degree
$N-1$. Thus, the largest spins of the bound states are $N-1$. This is
similar to the Regge behavior of hadronic resonances, albeit with an
imaginary slope. It remains
to be seen whether these poles are `physical' in the sense they
correspond to resonant states or as argued in ref.[1] are just
artifacts of our kinematical approximations. Nevertheless,
we will show
in the subsequent sections that the introduction of electromagnetism does
have an effect on their location in the complex $s$
plane.\footnote{\eightrm\baselineskip=10pt  For attempts in three and
four dimensions towards improving the `gravity eikonal' see [10-13] }
\vglue.4cm
\leftline{\tenbf 3.4 Charge-charge versus gravitational scattering}
\vglue.4cm

After having considered the pure gravitational scattering, we
introduce electromagnetic interactions. Once again we choose a frame such
that the `source' charge $e'$ has an electromagnetic shock wave
associated with it. The electric and magnetic fields on the shock
front are those found in the previous section, given in equation
(5). We assume at this point that the resultant effect of the combined
shock wave (gravitational and electromagnetic) on the test particle is to
produce a phase shift in its wave function which is the sum of
the individual phase shifts. This tacitly presumes the independence of
the gravitational and electromagnetic shock waves$^{2-4}$ which is by no
means self-evident and warrants justification. In a later section, we shall
present a first
attempt at such a justification$^{14}$. Since both the phase shifts are
proportional to $\ln {\mu}^2r_\perp^2$, the
net effect is succinctly captured by the shift $Gs \rightarrow Gs+ee'$, with
the final form of the wave function after it crosses the null plane
$x^{-}=0$ being $$
\psi _>(x^{\pm},x_{\perp})
{}~ =~\exp~\left[-i\left(ee^{'}+Gs\right)\ln\mu^{2}r^{2}_{\perp} +
ipx\right]~.
$$
Consequently, the scattering amplitude becomes
$$
f(s,t)= {k_+ \over 4\pi k_0}~{\delta(k_+ - p_+)}
{\Gamma (1-iee'-iGs) \over \Gamma (iee'+iGs)}\left (4 \over-t \right
)^{1-iee'-iGs}~~. \eqno(36) $$

\noindent
This gives the cross section,
$$
{d^2\sigma \over d\vec k_{\perp}^2}~ \sim~ {1
\over t^2} \left ( ee'
+ Gs \right )^2~~. \eqno(37)
$$
To compare the relative magnitudes of the two terms, we recall
that the electromagnetic coupling constant $ee'$ evolves only
with $t$ through radiative corrections and not with $s$. Thus in
the kinematical regime that we are considering, it remains fixed
at its low energy value. For example, if the particles
carry one electronic charge each, then $ee' \sim  1/137$.
On the other hand, at Planck scales, the second term in the cross
section is of order unity. This shows that gravity is the
principal contributor in the scattering process and
electromagnetic effects can be treated as small perturbations. Likewise,
the poles of the scattering amplitude (36) are shifted by ${\cal
O}(\alpha)$ corrections to the pure gravity poles. Observe that these
poles appear only when gravitational interactions are taken into account,
because it is only in this case that the interaction is a (monotonically
increasing) function of energy.

\vglue.4cm
\leftline{\tenbf 3.5 Charge-monopole versus gravitational scattering}
\vglue.4cm
Motivated by the conclusions of the last section, we now
proceed to investigate whether they undergo any modifications
when we assume one of the particles to carry a magnetic
charge. In other words, will gravity still dominate
over electromagnetic interactions at Planckian energies? With the
replacement of the electric charge $e'$ of the fast moving particle by a
magnetic charge $g$, the fields on the electromagnetic
shock front are given by equation (15). As before, when it crosses
the charge $e$, we add the
gravitational and electromagnetic phase shifts in its
wavefunction. While the former is still $-Gs\ln {r_{\perp}}^2$, the latter,
as seen from equation (2.39), is now $in\phi$. Thus, charge-monopole
electromagnetic effects cannot be incorporated by a shift of $Gs$, in
contrast to the charge-charge case. Thus the wavefunction assumes the form
$$ \psi _>(x^{\pm}, x_{\perp})~ =~
\exp~\left[i\left(n\phi - Gs\ln\mu^{2}r^{2}_{\perp} +
ipx \right) \right]
$$
Due to the azimuthal dependence, the calculation of the
overlap with momentum eigenstates has to be done ab initio.
Clearly, the relevant integral for the evaluation of
$f(s,t)$ is
$$\int d^2 r_{\bot}~\exp[i(n\phi - Gs~
\ln{ \mu ^2 r_{\bot}^2}  +\vec q . \vec r_{\bot})].$$
Once again, the integral over $\phi$ is readily done, and
the above reduces to
$$
{1 \over q^2} \int_0^{\infty} d \rho ~\rho^{1-2iGs} J_n(\rho)~.
$$
Here $J_n(\rho)$ is the Bessel function of order $n$. The
above integral is again a standard one$^4$ and
finally we get the amplitude
$$f(s,t)= {k_+ \over 4\pi k_0}~{\delta(k_+ - p_+)}~\left({n\over
2}-iGs\right){\Gamma
({n \over 2}-iGs) \over \Gamma ({n \over 2} + iGs)}\left (4 \over
-t \right
)^{1-iGs}~~ \eqno(38)
$$
and hence the cross section
$$
{d^2\sigma \over d\vec k_{\perp}^2}~ \sim~ {1 \over t^2} \left (
{n^2 \over4}
+ G^2s^2 \right )~~. \eqno(39)
$$
Since $n$ is at least of order unity,
it is clear from the above expression, that for $\sqrt s
\approx M_{pl}$, both the terms are of the same order of
magnitude. This means that unlike charge-charge scattering,
even at Planck scale gravity is no longer the dominant shock wave
interaction. Electromagnetism with monopoles becomes
equally important. This dramatic difference from the charge-charge case
is a consequence of the Dirac
quantization condition, which restricts the values of $e$
and $g$ from being arbitrarily small. In fact, the above may be
considered to be a
rephrasal of the strong coupling aspects of the monopole sector in
electromagnetism and of the gravitational interactions at Planck scale.
As already mentioned earlier, gravitational effects would indeed tend to
dominate for $Gs >>1$ if the Dirac quantum number $n$ is held fixed. But
it is far from clear if, in this circumstance, the simple-minded
semiclassical analysis performed above will go through without
modification. Indeed, as explained in ref.[3 , super-Planckian
energies will most probably entail real black hole collisions with the
ensuing technical complications.

Returning once more to the analytic structure of $f\left(s,t\right)$, we
see that now they occur at
$$Gs = -i\left( N + {n \over 2 } \right)~~, $$
that is a shift in $s$ by half-odd integral values. Once again, the
spectrum of these `bound states' is no longer a perturbation on the
spectrum in the pure gravity situation. More interestingly,
the shift observed above due primarily to the monopoles strongly suggest
another possibility: the {\it Saha phenomenon}$^{8}$. Recall that, this
implies that any charge-monopole pair composed of spinless particles
will, as a consequence of Dirac quantization, possess half-odd integral
quantized (field) angular momentum. If we blithely regard the integer $N$,
which also occurs in the spectrum of bound states in pure gravitational
scattering, as the {\it spin} of the states, then it is enticing to
consider the shift by one-half
the Dirac quantum number $n$ in the charge-monopole case to be the extra
spin due to the field angular momentum that the system would pick up in
accord with Saha's predictions. Further, if one speculatively associates
the Regge-like behavior observed
in purely gravitational scattering with the spectrum of some string
theory (albeit with imaginary slope parameter), then the spectrum with
charge-monopole electromagnetic scattering
can as well be speculated to correspond to some {\it supersymmetric}
string theory. In any event the role of electric-magnetic duality, were
we to actually discern any such string structures, can hardly be
over-emphasized.
\vglue.6cm
\leftline{\elevenbf 4. Shock Wave Mixing$^{14}$}
\vglue.4cm

We now turn to the issue of mixing of the two species of shock waves,
viz., gravitational and electromagnetic for the Planckian scattering of
charged particles. The gravitational shock wave relevant to two particle
scattering in
Minkowski space has been obtained in one of two ways$^9$ : either
by demanding that the Minkowskian geometry with a
lightlike particle present is the same as an empty Minkowski space with
coordinates shifted along the geodesic of the particle, or by a process
of `boosting' the metric of a massive particle to luminal velocities when
its mass exponentially decays to zero.  Since there are well known
(electrically and magnetically) charged black hole solutions of the
Einstein equation, boosting such solutions to the velocity of light would
of course produce both gravitational and electromagnetic shock waves. It
is then important to determine whether these two species of shock waves
actually {\it mix}. The
problem may be stated succinctly as follows: the calculation of the
amplitude in the shock wave picture
entails computing the phase factor that the shock wave induces on the
wave function of the target particle. When both particles carry charge,
phase factors are induced by electromagnetism and gravity
independently of the other. It has been assumed in the literature that ,
with both shock waves present, the net phase factor is simply the sum of
the individual phase factors$^{2-4}$. In other words, the gravitational and
electromagnetic shock waves are assumed to travel collinearly without
interaction, even though they are extremely localized singular field
configurations. We seek a possible justification of this supposition in
what follows.
\vglue.4cm
\leftline{\tenbf 4.1 Decoupling in the Reissner-Nordstr\"om Case}
\vglue.4cm

The gravitational field due to a stationary point particle of mass $M$ and
electric charge $Q$ is given by the standard Reissner-Nordstr\"om metric
$$ ds^2~=~~(1-{2GM \over r} + {GQ^2 \over r^2}) dt^2
{}~-~(1-{2GM \over r} + {GQ^2 \over r^2})^{-1} dr^2 ~-~r^2 d \Omega^2~,
\eqno(40)$$
where $G$ is Newton's constant. The question we address here is : if the
particle is Lorentz-boosted to a velocity $\beta \sim 1$, what will be
the nature of the gravitational field as observed in the `stationary'
frame? Let us assume, for simplicity and without loss of generality, that
the particle is boosted in the $+z$ direction, so that $z,t$ are related
to the transformed coordinates $Z,T$ according to
$$\eqalign{
T~&=~t\cosh \rho~ + ~z\sinh \rho,     \cr
Z~&=~t\sinh \rho~+~ z\cosh\rho~. } \eqno(41) $$
The parameter $\rho$ is called rapidity : $\beta=\tanh \rho$. The null
coordinates are defined as usual as $x^{\pm}=t \pm z$. The boosting
involves parametrizing the mass of the black hole as $M=2pe^{-\rho}$
where $p$ is the momentum of the boosted particle, lying almost entirely
in the longitudinal direction. $p$ is usually kept fixed at a large value
in the boosting process, and the limit of the boosted metric is evaluated
as $\rho \rightarrow \infty$. In this limit, the Reissner-Nordstr\"om
metric assumes the form
$$ ds^2~\rightarrow~dx^- \{ dx^+~-~dx^- [{2Gp \over |x^-|}~-~{GQ^2 \over
(x^-)^2} ] \}~-~d x_{\perp}^2~. \eqno(42) $$
The boosted metric does indeed seem to depend explicitly on the charge
$Q$. But, notice that this dependence is confined to a part of the
metric that can be removed by a diffeomorphism, albeit one that
is singular at the origin. However, the part that goes as $1/ |x^-|$
cannot be removed by any diffeomorphism; this latter, of course, is
precisely
the part that is associated with the gravitational shock wave$^3$.
Further,
not only is its coefficient independent of $Q$, it is identical to the
the coefficient of the $1/ |x^-|$ term in the boosted Schwarzschild
metric$^{3,9}$. All
memory of the charge of the parent black hole solution is obliterated
upon boosting, insofar as the gravitational shock wave is concerned.
Consequently, the mutual transparency of the two shock waves follows
immediately. \footnote{\eightrm \baselineskip=10pt We note here that our
approach and results for the
Reissner-Nordstr\"om case differ somewhat from those of ref. [15]
where, in fact, the limiting procedure employed appears to yield
vanishing electromagnetic shock waves in the luminal limit.} Hence the
net phase shift of the wave function of a test particle moving in the two
shock waves is simply the sum of the phase factors induced individually
by each shock wave. This, in the case of scattering of two electric
charges, simply amounts to the replacement $Gs \rightarrow Gs+ee'$ as
mentioned in ref.s [2,3]. A similar decoupling of
electromagnetic and gravitational shock waves can be seen by boosting a
{\it magnetically} charged Reissner-Nordstr\"om solution, which justifies
once again the determination of the net phase factor in the test charge
wave function as the sum of the individual phase factors in Planckian
charge-monopole scattering$^4$.

The foregoing analysis is completely general, and requires no assumption
on the strength of the charge $Q$, except perhaps that it be finite.
However, an analysis of the singularities of the the metric in (40)
indicates that, if one is to abide by the dictates of Cosmic Censorship,
the charge $Q$ must obey $Q \leq M$. In the extremal limit, the boosting
procedure adopted above forces $Q$ to decay exponentially to zero as
the rapidity runs to infinity. The electromagnetic shock wave, by
withering away in this limit, then trivially decouples from the
gravitational one. This was first pointed out in ref.[15].
\vglue.4cm
\leftline{\tenbf 4.2 Non-decoupling in Dilaton Gravity}
\vglue.4cm

The charged black hole solution of four dimensional dilaton gravity,
obtained as a part of an effective low energy theory from the heterotic
string compactified on some compact six-fold, is given, following
[16,17], as\footnote{\eightrm\baselineskip=10pt This metric is the
so-called string
metric$^{16}$. What follows is equally valid for the Einstein metric}
$$ ds^2~=~ (1-{\alpha \over Mr} )^{-1} \left [ (1 - {2GM \over r}) dt^2~-(1
-{2GM \over r})^{-1} dr^2~-~(1 -{\alpha \over Mr}) r^2 d \Omega^2
\right ]~~, \eqno(43) $$
where, $\alpha \equiv Q^2 e^{-2\phi_0} $, with $\phi_0$ being the
asymptotic value of the dilaton field $\phi$. The metric reduces to the
Schwarzschild metric when $\alpha=0$, and, not surprisingly, shares the
coordinate singularity at $r=2GM$ which becomes the event
horizon for the curvature singularity at $r=0$. In addition, there is
the `singularity' at $r={\alpha \over M}$ which is not necessarily a
coordinate singularity. We shall return to this point later.

We now apply the boosting procedure elaborated in the last section to
this metric. The mass of the black hole is parametrized as
$M=2pe^{-\rho}$ and the Lorentz-transformed metric is evaluated in the
limit as the rapidity $\rho \rightarrow \infty$ for fixed large $p$. The
result can be expressed as the Minkowski metric in terms of {\it shifted}
coordinate differentials,
$$ds^2~\rightarrow~d{\tilde x}^+ d{\tilde x}^-~-~(d{\tilde
x}_{\perp})^2~~, $$ where,
$$\eqalign{
d{\tilde x}^+ ~~&=~~dx^+ ~-~\left ( { {4Gp \over |x^-|} \over {1 - {\alpha
\over p |x^-|}}} \right ) dx^-  \cr
d{\tilde x}^-~~&=~~dx^- \left ( { {1 - {\alpha \over 2p |x^-|}} \over {1 -
{\alpha \over p |x^-|}} } \right )  \cr
d{\tilde {\vec x}}_{\perp}~~&=~~d{\vec x}_{\perp}~~. } \eqno(44) $$
Several features emerge immediately from these equations; of these, the
most striking is the explicit dependence on the charge $\alpha$ of terms
that will surely contribute to the gravitational shock wave because of
their non-differentiable functional form. No less important is the fact
that, in this case, the coordinate $x^-$ which, in the Schwarzschild (and
Reissner-Nordstr\"om) case(s), defined the null surface ($x^-=0$) along
which the two Minkowski spaces were to be glued, is now itself subject to
transformation by such a discontinuous function, again explicitly
depending on $\alpha$. Before examining these aspects in detail, we
note in passing that the results reduce to those in the Schwarzschild case
in the limit $\alpha=0$, as indeed is expected.

First of all, the charge $\alpha$ may be chosen to be small by taking a
large value of $\phi_0$, so that, with a large value of $p$, one can
binomially expand the denominators in the rhs of the first two equations
in (44); this yields, for points away from $x^-=0$,
$$ \eqalign{
d{\tilde x}^+~~&=~~dx^+ ~-~\left [{4Gp \over |x^-|} ~+~{4\alpha \over
(x^-)^2} \right ] dx^- ~+~ {\cal O}(\alpha ^2 / p)~  \cr
d{\tilde x}^-~~&=~~dx^-~+~{\alpha \over 2p |x^-|}~ dx^- ~+~ {\cal O}(
\alpha^2 / p^2)~. } \eqno(45) $$
We now observe that, as far as the shift in $dx^+$ is concerned, the part
that will contribute to the gravitational shock wave is in fact, {\it
independent} of $\alpha$, and, furthermore, is identical to the result in
the Schwarzschild and hence the Reissner-Nordstr\"om case. As for the latter
solution, the $\alpha$-dependent part may be rendered innocuous by a smooth
diffeomorphism.

This is however not the case for the shift in $dx^-$,
which is explicitly $\alpha$-dependent. Clearly, the gravitational shock
wave now possesses a more complicated geometrical structure than in the
earlier examples. The geometry can no longer be expressed as two
Minkowski spaces glued after a shift along the null surface $x^-=0$, for
now there is a {\it discontinuity} in the $x^-$ coordinate at that very
point, in contrast to the previous cases where it was continuous. This
discontinuity has a rather serious implication : unlike in the earlier
situation wherein the coordinate $x^-$ could well serve as the affine
parameter characterizing the null geodesic of a test particle crossing
the gravitational shock wave (cf. [3]), a null geodesic is
actually {\it incomplete} in this situation. To see this in more detail,
consider the geodesic equations  of a very light particle moving in the
Lorentz-boosted metric (42),
$$\eqalign{
{\dot T}~~&=~~\left({ {1- {\alpha \over Mr}} \over {1 - {2GM
\over r}} } \right ) E~~ \cr
r^2 {\dot \phi}~~&=~~\left ( 1 - {2GM \over r} \right ) L~~ \cr
{\dot r}^2~~&=~~(1- {\alpha \over Mr})^2 \left [ E^2~-~{L^2 \over r^2}
\left ({ {1 - {2GM \over r}} \over {1 - {\alpha \over Mr}} } \right )
\right ]~~. } \eqno(46) $$
Unlike the geodesic equations for a boosted Schwarzschild metric, which
can be solved perturbatively in a power series in the mass $M$ (or
alternatively in the parameter $e^{-\rho}$ (where $\rho$ is the
rapidity)$^9$, these equations do not admit any perturbative
solution because of the singularity at $r = \alpha/M$. Taking recourse to
singular perturbation theory does not evade the problem; the definition
of a {\it continuous} affine parameter is not possible in this case. It
follows that the singularity in question must be a {\it curvature}
singularity. Although for the Reissner-Nordstr\"om case also, for generic
value of the charge, the singularity at $r=0$ is no longer hidden by the
event horizon, there are no other singularities away from this point. In
the present instance, the singularity at $r=0$ is actually protected by
the Schwarzschild horizon. One might consider imposing an extremal
condition on the
charge $\alpha$ (vid. [16]): $\alpha = 2M^2$ to mitigate the
circumstances. However, this limit is not interesting for our
purpose, for the same reason that the extremal Reissner-Nordstr\"om is not
-- the charge decays exponentially to zero with the rapidity going off to
infinity.

The non-decoupling of gravitational and electromagnetic effects that we
see here can be made more articulate if one proceeds to actually
calculate the phase shift of the wave function of a test particle
encountering the gravitational shock wave, notwithstanding the pathologies
delineated above. The equations (44) above for the
differentials are consistent with the following finite shifts, obtained
by generalizing results of ref.[9],
$$\eqalign{
x^+ _{>}~~&=~~x^+ _{<}~+ ~2Gp \ln \mu^2 r_{\perp}^2~~ \cr
x^-_{>}~~&=~~x^-_{<}~+~{\alpha \over 2p} \ln \mu^2 r_{\perp}^2~~
\cr
{\vec x}_{>}~~&=~~{\vec x}_{<}~~. } \eqno(47) $$
With these, following [3] we can easily calculate the net phase
shift of the wave function of a test particle due purely to gravitational
effects:
$$  \Phi_{total}~~=~~(Gs~+~\alpha {k_{\perp}^2 \over 2s} ) \ln \mu^2
r_{\perp}^2~~. \eqno(48) $$
Here, $k_{\perp}$ is the transverse momentum of the test particle.
Thus, even if the test particle is electromagnetically neutral, its wave
function undergoes a phase shift that depends on the charge of the black
hole boosted to produce the gravitational shock wave. This is a novel
phenomenon, in our opinion, although, strictly speaking, in the
kinematical regime under consideration, the magnitude of the effect is
small. Nevertheless, the mixing of the electromagnetic and gravitational
shock waves, in this case is quite obvious.

The scattering amplitude for a test particle encountering such a
gravitational shock wave can be calculated following ref.[3].
Modulo standard kinematical factors and irrelevant constants, the answer
is
$$ f(s,t)~~\sim~~{1 \over t} { {\Gamma \left(1 - i(Gs + \alpha {k_{\perp}^2
\over 2s} ) \right )} \over {\Gamma \left(i (Gs + \alpha {k_{\perp}^2
\over 2s}) \right ) } }~~. \eqno(49) $$
Since the calculation is performed in a coordinate frame in which the
test particle is assumed to be moving slowly, the amplitude does not
appear manifestly Lorentz-invariant, although there is nowhere any
violation of Lorentz invariance . The only likely
outcome of such a calculation will be the replacement of the quantity
$k_{\perp}^2$ by the squared momentum transfer $t$ upto some numerical
coefficient of ${\cal O}(1)$. As a consequence, the poles in (49)
would undergo a shift of ${\cal O}(i \alpha Gt/N^2)$ from their
integer-valued (given by $N$) positions on the imaginary axis found in
the Schwarzschild case$^3$. This shift is quite different from
similar shifts when electromagnetic effects are included based upon a
decoupling assumption$^{1-3}$. The
non-decoupling is manifest from the coefficient $\alpha G$ in this case.
Also, the electromagnetic shifts are always constant independent of $t$,
in contrast to what we find here.
\vglue.6cm
\leftline{\elevenbf 5. Conclusions}

The predominance of shock waves as instantaneous mediators of
electromagnetic and gravitational interactions at Planckian
centre-of-mass energies and low fixed momentum transfers has thus been
established even for situations where a reduced local field theory
description is not immediately available. Charge-monopole interactions
show significant departures from ordinary charge-charge interactions at
Planckian energies to the extent that they begin to compete with
gravitational interactions at these energies.

The decoupling of electromagnetic and gravitational shock waves have now
been established for the case of general relativity, justifying thereby
earlier results incorporating both fields for electrically and
magnetically charged particles scattering at Planckian centre-of-mass
energies. The shifts in the poles due to electromagnetic effects stand
vindicated. Admittedly, in
the general relativity case the poles appear to be artifacts of the large
impact parameter approximation$^1$. However, as will be reported
elsewhere in the near future$^{13}$, the scattering amplitude does exhibit
poles even beyond the eikonal approximation. The nature of the
shift due to the dilaton coupling tends to reinforce the speculation
that string theory may actually provide a way to compute
corrections to this approximation as a power series in $t$.

The results may also have implications for black holes. The effect of
infalling
particles collapsing gravitationally onto a black hole has been
analyzed$^9$
to produce a shift of the classical event horizon. If we also
subscribe to the view$^{18}$ that this shift essentially involves
generalizing the flat space gravitational shock wave to a curved background,
then a particle whose fields are obtained by boosting fields of a dilaton
black hole would cause extra shifts of the horizon of a
Schwarzschild black hole. In addition, with electric and magnetic
charges present, novel contributions are to be expected for any
S-matrix proposal for dilatonic black holes.
\vfil /eject

\vglue0.6cm
\leftline{\tenbf 6. References}
\vglue0.4cm
\medskip
\itemitem{1.} H. Verlinde and E. Verlinde, {\it Nucl. Phys.} {\bf
B371}, 246 (1992); H. Verlinde and E. Verlinde, Princeton University
preprint PUPT-1319 (1993) (unpublished).

\itemitem{2.} R. Jackiw, D. Kabat and M. Ortiz, {\it Phys. Lett.} {\bf B
277}, 148 (1992).

\itemitem{3.} G.'t Hooft, {\it Phys. Lett.} {\bf B198} , 61 (1987);
{\it Nucl. Phys.} {\bf B304}, 867 (1988).

\itemitem{4.} S. Das and P. Majumdar, {\it Phys. Rev. Lett.} {\bf 72},
2524 (1994).

\itemitem{5.} A. S. Goldhaber, {\it Phys. Rev.} {\bf B140}, 1407 (1965).

\itemitem{6.} S. Das and P. Majumdar, IMSc-ICTP preprint IMSc/94-44,
IC/94/345, hep-th 9411058, to appear in {\it Phys. Rev. D}.

\itemitem{7.} J. Schwinger, {\it Science} {\bf 165}, 757 (1969).

\itemitem{8.} M. N. Saha, {\it Ind. J. Phys.}, {\bf 10}, 145 (1936);
{\it Phys. Rev.} {\bf 75}, 1968 (1949); H. A. Wilson, {\it Phys. Rev.}
{\bf 75}, 309 (1949).

\itemitem{9.} T. Dray and G.'t Hooft, {\it Nucl. Phys.} {\bf B253}
173 (1985).

\itemitem{10.} D. Amati, SISSA preprint SISSA-22-93-EP, and references
therein.

\itemitem{11.} M. Fabbrichesi, R. Pettorino, G. Veneziano and G.
Vilkoviskii, {\it Nucl. Phys.} {\bf B419}, 147 (1994).

\itemitem{12.} S. Deser, J. McCarthy and A. Steif, {\it Nucl. Phys.} {\bf
B412}, 305 (1994); see also M. Zeni, {\it Class. Quant. Grav.} {\bf 10},
905 (1993).

\itemitem{13.} S. Das and P. Majumdar, {\it Aspects of Planckian
Scattering Beyond the Eikonal}, IMSc preprint IMSc/94-95/61 (March 1995).

\itemitem{14.} S. Das and P. Majumdar, IMSc-ICTP preprint IMSc/94-95/46,
IC/94/365, hep-th 9411129 to appear in {\it Phys. Lett. B}.

\itemitem{15.} C. Loust\'o and N. S\'anchez, {\it Int. J. Mod. Phys.} {\bf
A5}, 915 (1990).

\itemitem{16.} G. Gibbons and K. Maeda, {\it Nucl. Phys.} {\bf B298}, 741
(1988).

\itemitem{17.} H. Garfinkle, G. Horowitz and A. Strominger, {\it Phys.
Rev. } {\bf D43}, 3140 (1991); {\it err.} {ibid.} {\bf D45}, 3888 (1992).

\itemitem{18.} G.'t Hooft, Nucl Phys, {\bf B335}, 138 (1990).

\vfill\eject

\bye